\PassOptionsToPackage{warn}{textcomp}
\documentclass[10pt, conference, compsocconf]{IEEEtran}\usepackage[]{graphicx}\usepackage[]{color}
\makeatletter
\def\maxwidth{ %
  \ifdim\Gin@nat@width>\linewidth
    \linewidth
  \else
    \Gin@nat@width
  \fi
}
\makeatother

\definecolor{fgcolor}{rgb}{0.345, 0.345, 0.345}

\usepackage{framed}
\makeatletter
\newenvironment{kframe}{%
 \def\at@end@of@kframe{}%
 \ifinner\ifhmode%
  \def\at@end@of@kframe{\end{minipage}}%
  \begin{minipage}{\columnwidth}%
 \fi\fi%
 \def\FrameCommand##1{\hskip\@totalleftmargin \hskip-\fboxsep
 \colorbox{shadecolor}{##1}\hskip-\fboxsep
     \hskip-\linewidth \hskip-\@totalleftmargin \hskip\columnwidth}%
 \MakeFramed {\advance\hsize-\width
   \@totalleftmargin\z@ \linewidth\hsize
   \@setminipage}}%
 {\par\unskip\endMakeFramed%
 \at@end@of@kframe}
\makeatother

\definecolor{shadecolor}{rgb}{.97, .97, .97}
\definecolor{messagecolor}{rgb}{0, 0, 0}
\definecolor{warningcolor}{rgb}{1, 0, 1}
\definecolor{errorcolor}{rgb}{1, 0, 0}
\newenvironment{knitrout}{}{} 

\usepackage{alltt}
\usepackage[utf8]{inputenc}
\usepackage{booktabs}
\usepackage[noadjust]{cite}
\usepackage{float}
\usepackage[hidelinks]{hyperref}
\usepackage{textcomp}
\usepackage{amsmath}
\usepackage{amssymb}
\usepackage{nccmath}
\usepackage{cuted}
\usepackage{dblfloatfix}
\usepackage{pifont}
\usepackage[table,x11names,dvipsnames,table]{xcolor}
\usepackage{tikz}
\usepackage{dblfloatfix}

\newcommand{\xmark}{\ding{55}}%
\ifCLASSINFOpdf
\graphicspath{{../img/}}
\else
\fi

\usepackage[]{color}
\usepackage{framed}

\makeatletter
\def\maxwidth{ %
  \ifdim\Gin@nat@width>\linewidth
    \linewidth
  \else
    \Gin@nat@width
  \fi
}
\makeatother

\definecolor{fgcolor}{rgb}{0.345, 0.345, 0.345}

\makeatletter
 {\par\unskip\endMakeFramed%
 \at@end@of@kframe}
\makeatother

\definecolor{shadecolor}{rgb}{.97, .97, .97}
\definecolor{messagecolor}{rgb}{0, 0, 0}
\definecolor{warningcolor}{rgb}{1, 0, 1}
\definecolor{errorcolor}{rgb}{1, 0, 0}

\renewenvironment{knitrout}{}{} 
\IfFileExists{upquote.sty}{\usepackage{upquote}}{}
\begin{document}
%
\title{How FAIR can you get?\\Image Retrieval as a Use Case to calculate FAIR Metrics}


\author{
    \IEEEauthorblockN{
        Tobias Weber,
        Dieter Kranzlmüller}
    \IEEEauthorblockA{
        \textit{Leibniz Supercomputing Centre}\\
        \textit{Bavarian Academy of Sciences and Humanities}\\
        Garching, Germany \\
        \{weber, kranzlmueller\}@lrz.de}
}

\IEEEoverridecommandlockouts
\IEEEpubid{\makebox[\columnwidth]{Publishing details to be added - accepted for IEEE escience 2018
\copyright2018
IEEE \hfill} \hspace{\columnsep}\makebox[\columnwidth]{ }} 

\maketitle

\begin{abstract}
A large number of services for research data management strive to adhere to the FAIR guiding principles for scientific data management and stewardship.
To evaluate these services and to indicate possible improvements,
use-case-centric metrics are needed as an addendum to existing metric frameworks.
The retrieval of spatially and temporally annotated images can exemplify such a use case.
The prototypical implementation indicates that currently no research data repository achieves the full score.
Suggestions on how to increase the score include automatic annotation based on the metadata inside the image file and support for content negotiation to retrieve the images.
These and other insights can lead to an improvement of data integration workflows, resulting in a better and more FAIR approach to manage research data.
\end{abstract}


%
\IEEEpeerreviewmaketitle

\section{Introduction}\label{sec:introduction}
All forms of digitised contents used as input for or output of scientific research activities are research data.
Metadata are data about these research data to make them Findable, Accessible, Interoperable, Reuseable (FAIR) amongst other features.
Research data products can be defined as the research data together with their metadata.
Scientific advances will be stimulated and the return of investment for research funding will increase, if research data are reused more often \cite{eosc}.
Enabling this reusage is a major challenge for researchers, since the necessary tasks are diverse, extensive, non-trivial, and often only recently added to their responsibilities \cite{bigDataLibraries}.
Many institutions, infrastructure projects and service providers support researchers with these data management tasks.

The FAIR guiding principles constitute quality criteria for research data products
and thus provide the necessary foundation to assess the services designed to help the researchers \cite{general003}.
Since they describe the requirements especially for scalable research data services,
they mainly focus on machine-to-machine interaction.

Recent proposals to derive specific metrics from these principles focus on single data sets \cite{oznome}\cite{metricspreprint}\cite{fairmetrics}.or on data from single research data repositories \cite{general024}.
While these metrics provide useful insights,
they do not focus on complex reusage scenarios across repositories.

Disambiguation of some FAIR principles is another problem that cannot be handled by current data-centric or repository-centric metrics:
Two central principles require, that data are richely described with "accurate and relevant attributes" \cite[principles F2 and R1]{general003}.
But the required richness does not have  a concise meaning unless the usage context is defined.
Since neither data items nor repositories can be used to derive this context in this respect cannot be measured so far.

This paper proposes a method to define use-case-centric metrics which do not share the two aforementioned drawbacks:
Use cases of data reusage across repositories give the necessary context to decide whether a data set is described richely enough and thus allow to derive a corresponding metric.
This approach adds to existing metric frameworks in a complementary way.


A case study to exemplify this method has been implemented:
The retrieval of spatially and temporally annotated images across research data repositories.
It is both relevant to many researchers and necessitates data integration over distributed sources.
For five research data repositories and 1.408.929 research data products the scores of the metric have been calculated.
The first calculation of the metric indicates that chronoreference and automatic accessibility are major issues.
The calculation can be executed automatically;
continuous executions will thus give evidence whether the first findings are robust and how the research data landscape changes over time.
Since our use case will be relevant to many, but not all disciplines and data-driven research methods,
other case studies can be implemented and reuse the methodological approach presented in this paper.

In \autoref{sec:related} the relation of this paper to other works is discussed.
\autoref{sec:methods} presents the rationale for the use case selection, states how a use-case-centric metric can generally be calculated and explains the methods used in developing the prototypical case study.
In \autoref{sec:results} an overview of the FAIR indicators collected is given.
\autoref{sec:discussion} discusses and evaluates the presented approach.

\section{Related Work}\label{sec:related}
Two data-centric metrics that can be used to measure FAIRness exist to our knowledge.
The metrics of \cite{oznome}, \cite{metricspreprint} can be applied to a whole repository by aggregating the measured values.

\cite{oznome} proposes a five star rating for research data alongside a tool for automatic assessment.
The authors of \cite{metricspreprint} introduce both a framework for measurable FAIRness of meta(data) and tools for semi-automatic assessment.
The framework allows to provide additional, possibly community-specific metrics.
Currently 14 examples for such metrics are described \cite{fairmetrics}.
Our approach can be integrated into the framework described in \cite{fairmetrics} and is fully automated.

The authors of \cite{general024} focus on measuring the FAIRness of research data repositories.
37 research data repositories have been manually assessed with a focus on Dutch and international providers.
The data to derive the repository-centric metrics is openly available \cite{datageneral024}.
Our results suggests a more critical perspective on accessibility compared to \cite{general024} - at least in the machine-actionable sense.

Relevant shortcomings  of data-centric and repository-centric metrics addressed by our approach are their inability to represent distributed retrieval scenarios and disambiguate some of the FAIR principles (see \autoref{sec:introduction}).
As far as we know, no other work tried to fill these gaps with use-case-centric metrics.

An architecture for FAIR-compliant research data integration across repositories is described in \cite{general025}.
This architecture has three main components:
The FAIR accessor is the first component (a Linked Data Platform Container), which consists of several MetaRecords (or FAIR profiles, the second component).
MetaRecords have themselves links to to FAIR projectors, the third component.
This allows for machine-actionable access to a data set or even a single data point in it.
The publication furthermore lists several community-specific efforts to realise research data integration that could be used to derive other use-case-centric metrics.
This approach can be mapped to our generic research data integration workflow (see \autoref{sec:methods}) and can therefore be extended to calculate another use-case-centric metric.
The architecture is based on semantic technologies and focuses primarily on data relevant for the life sciences.
The case study in this paper will use another type of technology and focus on a use case that is not specific to the life sciences.

The tools, standards and protocols presented in \cite{general026} are of high relevance for our use case since they present state of the art techniques in repository interoperability.
Image retrieval is a subset of information retrieval, one of the two major use cases of this primer.
Both of the major techniques we use in our implementation (OAI-PMH and Datacite) are listed in \cite{general026}. 

\cite{dataretrievalPreprint} provides an overview over practices in the retrieval of observational data across different disciplines.
The concepts found there provided valuable insights for the development of the generic research data integration workflow presented in \autoref{sec:methods}.
The review focuses more on manual information retrieval and discovery, but it could indicate other possible case studies, which then need to be automated.

The International Image Interoperability Framework\footnote{\url{http://iiif.io}} provides a convincing alternative to the set of protocols and services chosen by our implementation.
Whereas this project concentrates on pictures, the rationale behind the presented case study was extendability to other research data types and formats.

In \cite{oaipmh002} the usage of OAI-PMH for a search index and data discovery service is described.
Evaluation of the metadata formats presented in our paper or statistics comparable to ours are not included,
though some hints are given how different metadata formats can be used to realise the detection of data retrieval endpoints.

\cite{oaipmh001} discusses the difficulties in retrieving the data described by a metadata catalogue
provided via OAI-PMH.
Several metadata formats and their feature set are discussed, but not Datacite, the metadata format chosen for the presented implementation (Datacite didn't exist at the time \cite{oaipmh001} has been written).

The discussion of the first calculated metrics in \autoref{sec:results} adds to the overall description of the research data landscape given in \cite{general020}.
Whereas the statistical evaluation presented there discusses the broad range of research data repositories listed in re3data.org, we are more specific to data providers supporting our use case.

Research data retrieval can also be categorised as a big data integration task.
Further challenges and opportunities in managing big data have been described in many papers (e.g. \cite{relwork006}, \cite{general002}),
The five V's (volume, velocity, variety, value and veracity) of big data are e.g. presented in \cite{relwork010}.
Our case study has the management of variety in focus.
The proposed measurement of quality criteria can in principle be adapted to be applicable outside of the context of reuse of research data.

\section{Methods}\label{sec:methods}
\subsection{Selection of a Use Case}
In order to calculate use-case-centric metrics, a use case was needed,
which fulfills the requirements of relevance, complexity and effectiveness:
\begin{itemize}
    \item The use case has to be relevant for different fields of research.
    \item The use case has to include data reusage and integration across different data sources.
    \item The research data products that are integrated and collected due to the use case allow the reproducible calculation of a FAIR metric.
\end{itemize}
These criteria make sure that the use case is "interesting" enough while still being implementable.

As a use case satisfying these criteria in an optimum way, 
we chose the retrieval of images from research data repositories that are annotated with a certain place and time (of their creation).
Furthermore they must be licensed in a way that allows to determine whether scientific reusage is restricted in any way.

Image processing is a relevant technique for both the sciences and the humanities
(cf. for example \cite{imageExample001} and \cite{imageExample002}).
Selecting images on the basis of the date and location of their creation is generic enough to be still interesting for different disciplines and specific enough to be implementable.
Images reside in various different research data repositories - integration across different data sources is therefore an essential part of the use case.

As will be shown in the following subsection, FAIR metrics can be calculated alongside the implementation of this use case.
In the following, this calculation of FAIR metrics for our use case is called a case study.

\subsection{Calculation of Use-Case-Centric Metrics}
In this subsection the calculation of use-case-centric FAIR metrics (the case study) will be described in a generic fashion applicable to any use case.
A Description of our implementation will be given in the next subsection. 

Let $D$ be the set of all research data products of interest.
"Of interest" implies the non-consideration of research data products that are in principle not interesting for our use case.
The main criterion to separate data products of interest from the rest is their format:
An example is the non-consideration of tabulator-separated data for the image retrieval use case of this paper.
Calculating the proposed use-case-centric metric for these research data products would be pointless.

A central point for calculating use-case-centric FAIR metrics is the assessment of $n$ quality criteria that need to be met by a research data product $d \in D$ to be fully useable for the use case at hand.
Let $f_i$ : $D \rightarrow (0,1)$ ($1 \le i \le n$) be the assessment function which returns 1 when the criterion $i$ is met and 0 otherwise.
$d \in Q_i$ if and only if $f_i(d) = 1$.
$Q_i$ is hence the set of all data products meeting requirement $i$.

In the following paragraphs four use-case-centric FAIR scores will be presented, two for a single data product $d$ and two for a Research Data Repository (RDR) $R$.
Together they constitute the proposed metric.

\subsubsection{Fixed Score for Research Data Products}
For calculating this score, the weights of the use-case-specific quality criteria are fixed and evenly distributed:
\begin{equation}
    score_{\text{fixed}}(d) = \sum_{i=1}^{n} (f_i(d) \cdot \frac{1}{n})
\end{equation}
The range $score_{\text{fixed}}$ is the interval $[0,1]$,
with $score_{\text{fixed}}(d) = 1$ if $d$ meets all quality criteria.
Any value between 0 and 1 will indicate to what extent the research data product is useable for the use case.
If the case study to determine all $Q_i (1\le i \le n)$ is repeated, comparison of this score between two studies easily allows to measure trends.

\subsubsection{Relative Score for Research Data Products}

This score is added to give more importance to rarely met criteria.
To achieve this the weight of $i$ is calculated as a function of the size of $Q_i$ (which is the adoption rate at the time the case study is executed).
We define
\begin{equation}
    rareness(Q_i) = 1-\frac{|Q_i|}{|D|}
\end{equation}

and

\begin{equation}
    weight(Q_i) = \frac{rareness(Q_i)}{\sum_{j=1}^n{rareness(Q_j)}}
\end{equation}

The relative score for a research product $d$ is then calculated as follows:

\begin{equation}
    score_{\text{relative}}(d) = \sum_{i=1}^{n} f_i(d) \cdot weight(Q_i)
\end{equation}

\begin{figure*}[!b]
    \centering
    \includegraphics[width=\textwidth,clip]{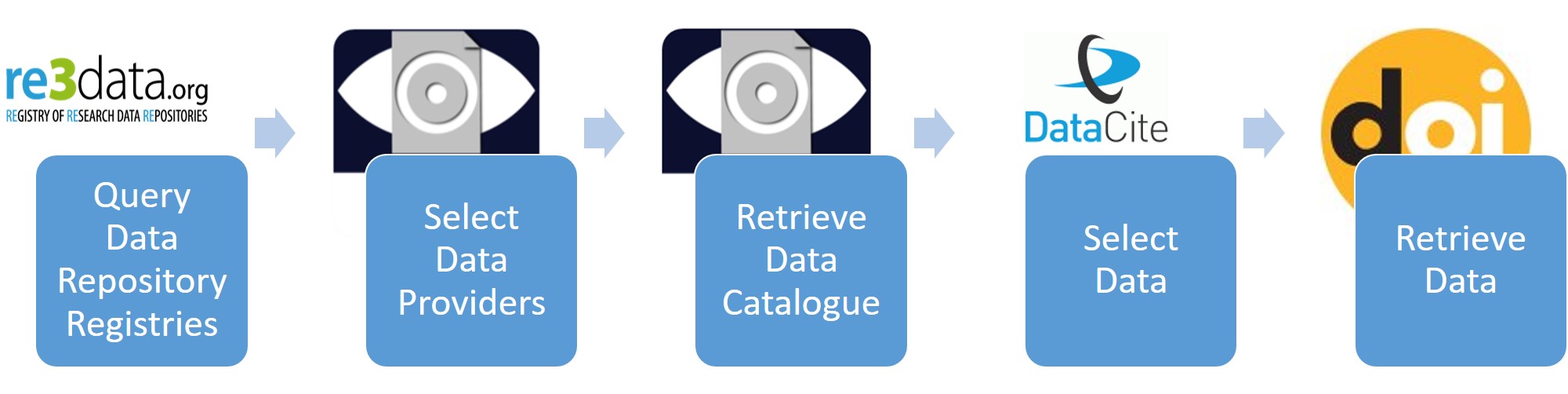}
    \caption{A Generic Workflow for Research Data Integration}
    \label{fig:genworkflow}
\end{figure*}
The range of $score_{\text{relative}}$ is still the interval $[0,1]$,
with $score_{\text{relative}}(d) = 1$ if $d$ meets all quality criteria;
but if two research data products $d_1$ and $d_2$ each meet one criterion, say $d_1 \in Q_1$ and $d_2 \in Q_2$, $d_1$ gets a higher score, if $|Q_1| < |Q_2|$.
Data products that implement rarely met criteria are hence highlighted positively, whereas criteria met by the major part of the data products have a smaller weight in this score.
When repeating the case study, it will turn out that $score_{\text{relative}}$ - in comparison to $score_{\text{fixed}}$ behaves somewhat more difficult to interpret:
it may be that $score_{\text{relative}}(d)$ decreases even though the number of met criteria for $d$ is constant between the two case study executions.
This is a consequence of the fact that weights are relative to the adoption rate, which is expected to be non-identical between case study repetitions.
Comparing two relative scores between two studies nevertheless gives an indication how the research product fares with regard to common standard.

The values of $weight(Q_i)$ of two or more case study executions can furthermore be compared to gain insights how the adoption rate of a certain criteria changed over time,
whereas the total sum of all $rareness(Q_i), (1\le i \le n)$ gives a hint how well-supported the use case is over all RDRs: the closer total rareness is to $n$, the less well-supported the use case is.

\subsubsection{Average Fixed Score for a Research Data Repository}
This metric is the arithmetic mean of fixed scores for data products $d$ managed in $R$.
Let $D_R$ be the set of all research data products of interest managed in $R$.

\begin{equation}
    score_{\text{avfixed}}(D_R) = \frac{\sum\limits_{d \in D_R} score_{\text{fixed}}(d)}{|D_R|}
\end{equation}

This metric can be interpreted on par with $score_{\text{fixed}}$:
Any value between 0 and 1 indicates to what percentage the research data products managed in $R$ are useable for the use case at hand with $score_{\text{avfixed}} = 1$ if $R$ only hosts full-quality research data products.
This number must be assessed together with the total number of research data products of interest $|D_R|$, since it is possible to get a score of 1 while hosting only one research data product of interest.

\subsubsection{Average Relative Score for a Research Data Repository}
This metric complements $score_{\text{avfixed}}$ as $score_{\text{relative}}$ complements $score_{\text{fixed}}$:

\begin{equation}
    score_{\text{avrelative}}(D_R) = \frac{\sum\limits_{d \in D_R} score_{\text{relative}}(d)}{|D_R|}
\end{equation}

To actually calculate these use-case-centric FAIR metrics for the chosen use case,
the quality criteria $Q_i$ need to be determined and the associated assessment functions $f_i$ need to be defined:
\begin{itemize}
    \item $d \in Q_{\text{chrono}}$ if and only if $d$ is annotated with the date when the image was taken.
    \item $d \in Q_{\text{geo}}$ if and only if $d$ is annotated with a reference to the location where the image was taken.
    \item $d \in Q_{\text{lic}}$ if and only if $d$ is annotated in a way that allows to determine whether $d$ may be used without any restrictions.
    \item $d \in Q_{\text{ret}}$ if and only if $d$ is automatically downloadable given only $d$'s metadata.
\end{itemize}

\subsection{Implementation}
The prototypical case study must implement the assessment functions $f_{\text{chrono}}, f_{\text{geo}}, f_{\text{lic}}$ and $f_{\text{ret}}$
alongside the use-case-specific data collection system for providing their input.
The implementation must furthermore fulfill the requirements of non-creativeness, automatability and repeatability:

\begin{enumerate}
    \item Only tools, standards and techniques that already exist may be used - implementation must be restricted to combining these. This is necessary since the case study should measure the status quo, not improve it.
    \item The integration must not include manual effort.
        This means it is 'machine-actionable' (for an explanation see \cite{general003}).
        This requirement ensures that our implementation scales with the number of research data products and repositories and that it is insightful with regard to automated data integration workflows.
    \item The case study execution has to be repeatable to ensure the ability to measure trends. This requirement furthermore guarantees reproducibility of the measured results.
\end{enumerate}

The implementation and hence this subsection is organised along the five steps of a generic research data integration workflow as depicted in \autoref{fig:genworkflow}.\footnote{All
    source code (DOI \href{https://doi.org/10.25927/001}{10.25927/001})
    and data (DOI \href{https://doi.org/10.25927/000}{10.25927/000})
    necessary to reproduce or replicate our findings are published as accompanying resources to this publication.
}
It is compatible to a generic research data infrastructure as described in \cite{general022}.
Any use case of research data integration should roughly be mappable to this generic workflow.

Once the use case of data integration has been implemented, the implementation of the case study is taken care of, centering around the steps to calculate the FAIR metrics.

All steps are implemented using the python programming language.\footnote{\url{https://www.python.org}}
Shell scripts wrap the python scripts to make sure the scripts are called in the correct order and with consistent parametrisation 
(the shell scripts also orchestrate the parallelisation).
All processed data necessary to replicate our findings are made available.
For each step technical options will be discussed and the selected implementation of the use case and the case study will be sketched.

\subsubsection{Querying Data Provider Registries}

The first step of the generic workflow (\autoref{fig:genworkflow}, left-hand side) consists of three tasks: Compiling a list of data providers, assessing the quality of data management of these providers and creating a list of APIs supported by them.
Assessing the quality of the data services of a data provider heavily depends on personal interaction:
It includes finding and reviewing relevant policies, certification audits, and last but not least, experience in using the provided services.
Suitable registries must provide this quality assessment information via an API.

According to our research, there is only one candidate fulfilling the requirements to a research data provider registry at the moment:
The Registry of Research Data Repositories (re3data.org).\footnote{\url{https://www.re3data.org}}
The number of repositories registered in re3data.org grows steadily:
From 400 repositories listed in July 2013 \cite{general019},
it more than quintupled to 2093 repositories by June 2018.
re3data.org also lists data providers that are not research institutions in the classical sense, but nevertheless provide valuable data for researchers
such as the Climate Data Centre\footnote{\url{https://cdc.dwd.de}}
or national statistic agencies\footnote{e.g.: \url{https://www.ons.gov.uk}}.
Another registry, Databib (databib.org) has been integrated into re3data.org, \cite{databib}.
The Directory of Open Access Repositories (OpenDOAR)\footnote{\url{http://www.opendoar.org/}} and the Registry of Open Access Repositories (ROAR)\footnote{\url{http://roar.eprints.org/}} have also been evaluated.
Both provide a means to automatically retrieve a list of repositories, including API endpoints (exclusively OAI-PMH).
Since they are both primarily focused on open access publication of scholarly articles and do not include quality assessment information,

they don't meet the requirements of the case study implementation.

The output of the implemented queries to the re3data.org-API is a list of RDRs including information about their supported APIs and information about their quality management.
The raw output is then processed and the relevant information is stored in a Comma Separated Values file (CSV) describing the RDRs.
Additional information offered by re3data.org (e.g. about licenses and formats) will not be used,
since the way the scores are calculated requires that we retrieve them on a data item basis and not aggregated over all data provided by a repository.

\subsubsection{Selection of suitable data providers}

Decisive criteria in the selection of suitable RDRs are the feature set supported by their APIs and the metadata format they use to describe their data items.
In the following both criteria will be discussed separately.

The \textbf{APIs} will be used for three tasks:
to retrieve the information which metadata formats are supported, to get the information how many data items are hosted by the RDR and to retrieve the data catalogue (i.e. all the metadata, see step three).

The API which may satisfy our requirements are listed in \autoref{tab:methods} together with the numbers of RDRs providing the API and the adoption rate as given by re3data.org and retrieved during step one.
We left out some listed APIs on purpose since their design rationale is not compatible for our requirements.
{1165 RDRs (56\% of them) have no API at all registered.

\begin{knitrout}
\definecolor{shadecolor}{rgb}{0.969, 0.969, 0.969}\color{fgcolor}\rowcolors{2}{gray!6}{white}
\begin{table}

\caption{\label{tab:methods}APIs supported by RDRs listed in re3data.org}
\centering
\begin{tabular}[t]{lrr}
\hiderowcolors
\toprule
API & Absolute & Relative\\
\midrule
\showrowcolors
REST & 304 & 14.52 \%\\
OAI-PMH & 162 & 7.74 \%\\
SOAP & 68 & 3.25 \%\\
SPARQL & 27 & 1.29 \%\\
\bottomrule
\multicolumn{3}{l}{\textbf{Note: }  n = 2093}\\
\end{tabular}
\end{table}
\rowcolors{2}{white}{white}

\end{knitrout}

Representational State Transfer APIs (ReST),
the Simple Object Access Protocol (SOAP) \cite{SOAP},
and the SPARQL Protocol and RDF Query Language (SPARQL) \cite{SPARQL}
are less viable options for our use case:
While each API endpoint of a certain type could support the generic workflow of research data integration,
each does it in a different way:
Whereas one ReST-API might provide access to "collections", the other one has "datasets" as the basic data type.
This makes it very cumbersome to utilise the APIs across RDRs,
for example to retrieve the data catalogue (see next step) in a uniform way.
Notwithstanding proposals to achieve the necessary level of homogeneity \cite{heteroAPIs},
we could only find one example where this has been achieved for SOAP in the context of cancer research \cite{soapintegration}.
Only extensive implementation work could allow for an uniform access across RDRs.
In the context of the present work this effort would violate the non-creativeness and the automatability requirement.

The Open Archive Initiative Protocol for Metadata Harvesting (OAI-PMH)\footnote{\url{http://www.openarchives.org/OAI/openarchivesprotocol.html}}
has an adoption rate big enough to be relevant and supports data catalogue retrieval.
There is a positive trend since 2015 for OAI-PMH: both the absolute number of RDRs supporting it and the adoption rate increased (in 2015 there were 85 RDRs, which entails an adoption rate of 6.2 \%, cp. \cite{general020}).
OAI-PMH provides the necessary semantics to uniformly implement step two and three of both the use case and case study.
The protocol is embedded into HTTP and supports several operations to retrieve information about a RDR and the research data products managed in it:
\begin{itemize}
    \item ListMetadataFormats: returns a list of metadata formats.
    \item ListRecords: returns a list of metadata records describing data items.
\end{itemize}

The prototypical implementation presented here is therefore based on OAI-PMH.

Suitable \textbf{metadata formats} provide the necessary information to implement the accessor functions $f_{\text{chrono}}$, $f_{\text{geo}}$, $f_{\text{lic}}$ and $f_{\text{ret}}$ and they need to indicate whether the described data item is an image and therefore of interest for the use case.
The 49 metadata formats that are offered by the RDRs providing an OAI-PMH interface have been evaluated alongside the following three criteria
(If a metadata format is available in multiple versions, the most expressive one has been evaluated):
\begin{itemize}
    \item Existence: There are predefined fields in the format which has the necessary semantics to implement $f_{\text{chrono}}$, $f_{\text{geo}}$ and $f_{\text{lic}}$
        $f_{\text{ret}}$ is not implementable with the metadata alone, but there need to be sufficient information to determine the protocol and endpoint for data retrieval.
        Furthermore, there has to be a field indicating whether the described data are research data of interest, i.e. a field for the data type.
    \item Unambiguity: The information is placed at a uniquely defined place in the format and can be used as is (i.e. without case distinction, additional retrieval, e.g. of ontologies).
    \item The fields are mandatory i.e. we can assume they are always part of a valid metadata record (this is an optional requirement).
\end{itemize}

Datacite is the most widespread standard fulfilling all requirements when we sort the metadata formats in the decreasing order of number of RDRs supporting it:
With 10 RDRs supporting it,
Datacite has been adopted by 10.87 \% of the RDRs providing an OAI-PMH interface.
In fact, Datacite is more relevant than these numbers suggest, since the RDRs supporting it are among the biggest RDRs in terms of the number of research data products (see \autoref{sec:results}).

With qualified elements for date, license and georeference Datacite provides the necessary input to implement the accessor functions.
Additionally Datacite's identifier field is mandatory and has to contain a Digital Object Identifier (DOI)\footnote{\url{https://www.doi.org}}.
Since DOIs can be resolved to URLs to retrieve the data item, an implementation of $f_{\text{ret}}$ is possible.
This features stands out in comparison with other metadata formats.
Datacite supports two elements to detect whether the metadata describe images:
On the one hand the "generalResourceType"-field, with "Image" as a possible value (these are determined by a controlled vocabulary),
on the other hand the "formats"-element of Datacite.
This element can specify the mime-type of complex research data,
which can be used here to include those data sets whose formats match the "image/*" pattern.
Datacite will therefore be the metadata format of choice for the implementation.

With all these choices in mind, step two ("selection of suitable data providers") is implemented as an iteration of all RDRs selected in step one, and a ListMetadataFormats-OAI-PMH query to filter out non-functional RDRs and those which do not support Datacite.
The output from the previous step is updated accordingly.

\begin{knitrout}
\definecolor{shadecolor}{rgb}{0.969, 0.969, 0.969}\color{fgcolor}\rowcolors{2}{gray!6}{white}
\begin{table*}

\caption{\label{tab:rdr_score}Score of RDRs}
\centering
\begin{tabular}[t]{lrrrrrrr}
\hiderowcolors
\toprule
RDR name & Items of Interest & $score_{\text{avfixed}}$ & $score_{\text{avrelative}}$ & $f_{\text{chrono}} = 1$ & $f_{\text{geo}} = 1$ & $f_{\text{lic}} = 1$ & $f_{\text{ret}} = 1$\\
\midrule
\showrowcolors
figshare & 1.224.071 & 0.0000004 & 0.0000004 & 0 & 0 & 0 & 2\\
Zenodo & 184.796 & 0.2500000 & 0.2245688 & 0 & 0 & 184.796 & 0\\
PANGAEA & 35 & 0.6642857 & 0.6558059 & 0 & 29 & 32 & 32\\
PUB Data Publications & 18 & 0.2500000 & 0.2245688 & 0 & 0 & 18 & 0\\
GFZ Data Services & 9 & 0.5277778 & 0.5230702 & 8 & 5 & 6 & 0\\
\bottomrule
\multicolumn{8}{l}{\textbf{Note: }  n = 1.408.929}\\
\end{tabular}
\end{table*}
\rowcolors{2}{white}{white}

\end{knitrout}
\subsubsection{Retrieval of a Data Catalogue}

The retrieval of the data catalogue acts as a necessary precondition to select data of interest and to assess the metadata with the accessor functions.
During this step all RDRs providing metadata in a Datacite format are harvested via the OAI-PMH protocol.
The harvesting has been distributed to four parallel working harvester processes.
Any number of harvest workers can be used, which is a valuable option, should the number of RDRs increase, which are supporting the technology of our use case.
The code allows us to balance the harvest of the RDRs (allocating special resources to bigger RDRs and bundling smaller RDRs together).

Often, a trade-off has to be made: On the one hand investing too much in handling errors caused by missing compliance with the OAI-PMH standard or poor service quality would violate the non-creativeness requirement.
On the other hand getting more metadata records was necessary to calculate meaningful metrics.
To name but one example: During the harvest the timeout for an HTTP request has been set to 20 seconds and the harvester has even been programmed to retry once more after a socket timeout.
Nevertheless some RDRs couldn't be harvested to the full extent.

The output of the step three is a distributed data catalogue in the Datacite format of all RDRs selected in step two.

\subsubsection{Data Selection}
Considering the use case, this step consists of the processing of the retrieved data catalogues,
its merging and filtering out data items that are not of interest.
For the use case, this would lead to a subset of a merged data catalogue,
only comprising records for research data products that match certain search criteria (e.g. images annotated with a specific time of creation or license).
For the case study, in contract, all research data products are enlisted with the specific values and evaluated with the accessor functions:

\begin{itemize}
    \item $f_{\text{lic}}$ checks whether the attribute "rightsURI" is filled at least once with a valid URL. Only this field allows for a unambiguous identification of the effective license (the rights field itself allows free text).
    \item $f_{\text{geo}}$ checks whether the geoLocations element has at least one child which contents validates.
    \item $f_{\text{chrono}}$ checks whether the dates element has a child with dateType attribute set to "Created".
\end{itemize}

Step four has been executed on twelve worker processes, but the code allows scaling to any number.
Output of this step is a list for each data item of interest including the corresponding accessor function result and some additional administrative information.

\subsubsection{Data Retrieval}
The fifth step of the generic workflow of research data integration is the retrieval of the data based on the list of data items defined in the previous step.
The harmonisation of the data is out of scope of our use case and case study.

For five RDRs the data catalogue included data of interest in the sense defined in \autoref{sec:methods}.
These five are not homogeneous in their support for data retrieval.
None of the APIs registered at re3data.org offered a uniform and automatable way to determine the protocol and endpoint for data retrieval based on the metadata alone.

Since DOIs are mandatory for Datacite metadata, resolving the DOI to a URL results at least in a request to a human-readable page (a so-called splash or landing page).
But these pages provide no machine-actionable retrieval mechanism: a human being can identify a download link, if it is present on the page, whereas a crawler only sees a number of unqualified links.
Screen-scraping all landing pages would violate all three requirements for the prototype.

The only machine-actionable option therefore is to use HTTP Content negotiation \cite{RFC7230} to get the access to the images, which did work for research data products from two RDRs (Pangaea and figshare).
Content negotiation allows to systematically retrieve the data, in principle even if the protocol has to change (e.g. from HTTP to FTP).
Content negotiation also allows to retrieve different representations of the research data products (e.g. bibtex-formatted citations of the research data products).

$f_{\text{ret}}$ can therefore be implemented on the basis of content negotiation:
The DOI URL is requested via HTTP with the additional header set as "Accept: image/*". If the request results in 200 HTTP status code eventually (iterating over all redirecting HTTP status codes) and a Content-type header field is set that matches the image/* wildcard, $f_{\text{ret}}$ returns 1.
If this client-initiated content negotiation fails, server-sided content negotiation can be probed:
If the server sets a HTTP Link header (see \cite{RFC5988}) in its last reply, the value of the header can include the necessary data retrieval information:
If one of the link-values has a type-field set to a mime-type that is identical to a annotated format in the metadata, the URI-reference is requested.
If the request results in a 200 HTTP status code eventually (again after following possible redirects) and a Content-type that matches this specific mime-type, $f_{\text{ret}}$ returns 1.
In all other cases $f_{\text{ret}}$ returns 0.

If we assume that the check for $f_{\text{ret}}$ takes about two seconds on average a sequential calculation for 1,4 million records would take more than 32 days.
Parallelisation is therefore a necessity.
With 34 retrieval workers checking $f_{\text{ret}}$ in parallel, step 5 took less than one day.
The number or retrieval workers can again scale, if the number of records to be checked might increase in the future.

\begin{knitrout}
\definecolor{shadecolor}{rgb}{0.969, 0.969, 0.969}\color{fgcolor}\rowcolors{2}{gray!6}{white}
\begin{table}

\caption{\label{tab:qc}Rareness and Weight for Quality Criteria}
\centering
\begin{tabular}[t]{lrrr}
\hiderowcolors
\toprule
Quality Criteria & $|Q_i|$ & Rareness & Weight\\
\midrule
\showrowcolors
$Q_{ lic }$ & 184852 & 0.8687996 & 0.2245688\\
$Q_{ geo }$ & 34 & 0.9999759 & 0.2584755\\
$Q_{ ret }$ & 34 & 0.9999759 & 0.2584755\\
$Q_{ chrono }$ & 8 & 0.9999943 & 0.2584802\\
\bottomrule
\multicolumn{4}{l}{\textbf{Note: }  n = 1.408.929 - total rareness = 3.87}\\
\end{tabular}
\end{table}
\rowcolors{2}{white}{white}

\end{knitrout}

\begin{knitrout}
\definecolor{shadecolor}{rgb}{0.969, 0.969, 0.969}\color{fgcolor}\rowcolors{2}{gray!6}{white}
\begin{table*}[!b]
\caption{\label{tab:fair}Case study coverage of FAIR principles}
\centering
\begin{tabular}[!b]{lrr}
\hiderowcolors
\toprule
Principle & Coverage\\
\midrule
\showrowcolors
F1 "(meta)data are assigned a globally unique and persistent identifier" & \checkmark \\
    F2 "data are described with rich metadata" & $Q_{\text{geo}}$, $Q_{\text{chrono}}$ \\
F3 "metadata clearly and explicitly include the identifier of the data it describes" & \checkmark \\
F4 "(meta)data are registered or indexed in a searchable resource" &  \checkmark\\
\midrule
    A1 "(meta)data are retrievable by their identifier using a standardized communications protocol" & $Q_{\text{ret}}$\\
    A1.1 "the protocol is open, free, and universally implementable" & $Q_{\text{ret}}$\\
    A1.2 "the protocol allows for an authentication and authorization procedure, where necessary" & $Q_{\text{ret}}$\\
A2 "metadata are accessible, even when the data are no longer available" & \xmark\\
\midrule
I1 "(meta)data use a formal, accessible, shared, and broadly applicable language for knowledge representation." & \checkmark\\
I2 "(meta)data use vocabularies that follow FAIR principles" & \xmark\\
I3 (meta)data include qualified references to other (meta)data & \xmark\\
\midrule
R1 "meta(data) are richely described with a plurality of accurate and relevant attributes" &  $Q_{\text{geo}}$, $Q_{\text{chrono}}$ \\
    R1.1 "(meta)data are released with a clear and accessible data usage license" & $Q_{\text{lic}}$ \\
R1.2 "(meta)data are associated with detailed provenance" & \xmark \\
R1.3 "(meta)data meet domain-relevant community standards" & \xmark \\
\bottomrule
\end{tabular}
\end{table*}
\rowcolors{2}{white}{white}
\end{knitrout}
\section{Results}\label{sec:results}

In total we retrieved Datacite-metadata for 1.408.929 images, i.e. research data products of interest.
The numbers aggregated over the RDRs are shown in \autoref{tab:rdr_score}.
Since these data rely on only one run in June 2018, they need to be interpreted with care
(cf. the corresponding considerations in \autoref{subsec:limitations}).
Nevertheless, a first preliminary interpretation will be given in the following paragraphs.
This interpretation is up to revision when the case study has been executed more often.

With regard to the calculation of the FAIR metrics the method proposed in \autoref{sec:methods} leads to reasonable results.
Since the images managed in Pangaea and the GFZ Data Services support three out of four quality criteria almost completely, their score is high.
The relatively low compliance of images managed in figshare manifests in small scores.

The most surprising result of the first case study execution is the low number of temporal annotations.
It was to be expected that existing metadata in the image files would lead to a high number of annotations:
Many formats can be automatically analysed with open source tools, like the exif tool suite.\footnote{see \url{https://www.sno.phy.queensu.ca/~phil/exiftool/}}
This possibility has been checked with random samples out of the images with the mime type "image/jpeg".
In all cases this turned out to be a viable option to automatically determine the date of creation of the image.

Similar considerations apply to georeferential annotations, but to a lesser extent:
Whether geo-tagging of images is part of the image-file's metadata is dependent on the feature set supported by the device used for its creation (date-tagging should be more common).

The number of data items that are annotated with a license-URI is large;
hence it brings out the difference between the fixed and relative scores.
Only license information that is given by a URL are accepted,
which is the reason why $|Q_{\text{lic}}|$ is lower than could be expected (the free-text license field has been ignored).

The value of $|Q_{\text{ret}}|$ has been expected to be low, since automatic retrievability of heterogeneous research data is a challenging task.
The first case study execution indicates that this assumption is justified.

According to the data collection no single research data product satisfies all quality criteria.
Some steps to improve this situation are proposed in the following section.

The calculated rareness and weight numbers can be read off from \autoref{tab:qc}.
They are unproportionally affected by the size of the two biggest RDRs, figshare and Zenodo.
The total rareness is the sum of all rareness values.
Its value (3.87) is very close to the maximum (4, or number of quality criteria).
This value can be taken as a general indicator how well the set of all tested RDRs supports the use case and its tech stack.

\section{Discussion}\label{sec:discussion}
The prototypical implementation of the use-case-centric FAIR metrics presented here accounts for 10 out of the 15 FAIR principles \cite{general003}, as can bee seen in \autoref{tab:fair}.
A checkmark depicts principles that are covered by the overall execution and not by a specific quality criteria.
The main claim of the presented approach consists in the hypothesis that the use case chosen provides a specific meaning to the vague criteria F2 and R1 and therefore makes adherence to these principles measurable.
The implementation is a proof-of-concept to justify this claim.

F1, F3 and A1 are accounted for, since Datacite requires DOIs as a research data identifier.
Since only HTTP and open and freely available protocols based on HTTP are used in the case study A1.1 is covered.
F4 is covered, since re3data.org is queried as a searchable resource.
To check authentication or authorisation is not part of the case study, but A1.2 is nevertheless covered, since HTTP allows for this functionality.
Metadata in Datacite format are taken to be a language for knowledge representation, therefore the score also checks for I1.

R.1.3 is not the focus of the generic use case chosen for the implementation.
Nevertheless the checked quality criteria of the use case could be extended to cover domain-relevant community standards,
hence disambiguating another vague criterion.

A query to the "Bielefeld Academic Search Engine" (BASE)\footnote{\url{https://www.base-search.net}} \cite{relwork002} in June 2018 with comparable search parameters (restricting to type "Still image") resulted in about 8,5 million results.
This indicates that the first run of the prototype covered an essential part of the available images in the context of research data.
BASE includes more than 129 million records in June 2018, primarily focusing on publications (which includes but is not limited to research data as understood in this paper).

\subsection{Caveats and limits of validity}\label{subsec:limitations}
Although our use case is quite generic and should therefore capture many aspects of research data integration, it will definitely not capture all.
There will be disciplines that find another use case or another technology better-suited for a use-case-centric metric.
While their implementation and set of all quality criteria $Q_i (1\le i \le n)$ will hence differ from the one presented in this paper as a proof of concept, the general approach to calculate the metric will still apply.

Although the greatest care was taken in its implementation, the case study is software and hence bugs are a possibility.
A stricter implementation which has fewer workarounds or an implementation with an additional set of features or another technique on the base level of the implementation might lead to differing metrics for a specific research data product or RDR.
Since this is always the case if software is used to produce scientific data, we hence publish the source code of the study with a license, that allows reusage and modifications,
as long as the resulting code is itself made publicly available.
This allows for code adaption and reruns of the metric to reproduce or refute our measurements.
The idea of use-case-driven FAIR metrics is nevertheless not invalidated by these kinds of shortcomings of prototypical implementations.

A case-study-execution is in principle time-biased.
This feature of use-case-centric metrics is independent of the implementation, unlike the aforementioned aspects:
Some services could have maintenance activities during the execution.
As a first countermeasure we repeated unsuccessful harvests to handle this threat to validity.
The execution of a case study should always be repeated on a regular basis and the results should always be annotated with the date of the case study execution.
This handling will help to detect and treat shortcomings due to the time bias.

Since the overall process to collect the data and calculate the metric currently takes about two and a half days, it is not feasible to rerun it too often to get a bigger sample size.
Another problem in this context is the load step three and five causes on the RDRs,
which could be taken for an attack if they run "too efficiently". 
The regular repetition is planned in the future.
A reasonable schedule would be to execute the case study twice a month.

\subsection{Lessons learned}\label{subsec:gap}
In this subsection lessons learned from the use case implementation are given.
These are directed primarily at two target groups:
\begin{itemize}
    \item \emph{RDR providers}: Maintainers of RDRs could be interested in improving their score or in enabling their RDR to be evaluated in a case study like the presented one.
    \item \emph{Researchers}: They could be interested in an implementation of a case study to support another use case or they could want to implement a workflow for integration of research data based on the technology used (with or without the intent of measuring FAIRness).
\end{itemize}

Providers of RDRs should consider registering to re3data.org and offering OAI-PMH as an API to increase the impact of their data.
re3data.org is a valuable asset to automate research data integration tasks across RDRs.
Some information are inconsistent (e.g. some URLs of APIs registered on re3data.org are not pointing to the actual API endpoint, but to the documentation) or out-of-date.
RDR maintainers should regularly check them and request an update.
To guarantee interoperability of registry retrievals as in step one and two of our generic workflow, future services other than re3data.org should be compliant to their schema \cite{re3dataschema}, or at least provide a mapping.

OAI-PMH is currently the best option for providers of RDRs, if they want to support the generic workflow of research data integration (cf. \autoref{fig:genworkflow}).
Nevertheless there are reasons for an update (current version is 2.0) or an alternative protocol supporting the semantics sketched in \autoref{sec:methods}:
Improvements for a retrieval of a complete data catalogue could be achieved if more efficient protocols would be supported (e.g. a compressed metadata dump retrievable by FTP which is in the format of an unchunked ListRecords or ListIdentifiers response).
Other missing features are the possibility to retrieve the number of records available via a specific metadata format
or a way to retrieve the size of the chunks (how many metadata records are returned per request).

Providers of RDRs which already offer an OAI-PMH interface should review its compliance with the standard.
56 or 34.57\% out of 162 RDRs have an unresponsive or uncompliant OAI-PMH endpoint.
The usage of resumption tokens to chunk the harvest into manageable packages should be considered if the RDR has more than 1000 items or supports potentially extensive metadata formats, such as Datacite.
The size of the chunks provided by the OAI-PMH server should also be chosen with consideration.
If the chunk size includes only 10 records per request and the OAI-PMH interface is limited to one request per second, step three of our implementation becomes very time-consuming.
Under these circumstances the harvest of a RDR with over two million records would take more than two days (in the best case).

Image formats such as jpeg support automatic annotation by providing the metadata as a part of the file format.
Automatic annotation procedures on ingest or by iterating over the existing research data stock could be implemented to increase the number of temporally (and probably also spatially) annotated data.
This will strongly increase the score of the research data products.

Researchers and data stewards should always use URIs to identify the license of the research data. Free text does not allow for a doubtless and machine-actionable determination under which conditions a research data item can be reused.

Considering the retrieval of data items more RDRs should follow the example set by PANGAEA and use the Link header to enable direct resource downloading or alternatively honour client-side content negotiation.

To complete the picture with regard to the FAIRness of the research data landscape, additional case studies should be designed, executed and their results published.
As already stated in \autoref{sec:related}, \cite{general025} might be a good option.
The research data integration technology presented in \cite{general025} is based on another technology (semantic web and linked data)and its architecture is compatible with the presented approach to calculate use-case-centric FAIR metrics.

Another use case could include the retrieval of statistical data in a text-based content type (such as CSV files).
Finding data sets fitting a specific research question and automatically harmonising and evaluating the full data set is a non-trivial but common task.

A third possibility would be to find a use case that relies on the five FAIR guiding principles which were not covered by our case study.

\section{Conclusion}\label{sec:conclusion}

In this paper an approach to design and implement use-case-centric FAIR metrics has been presented.
This approach allows to cover research data integration across different data sources in the calculated metric.
The use case furthermore provides the necessary context to disambiguate two of the FAIR guiding principles, which have not been measurable so far.
The prototypical implementation shows the viability of this approach.
The method and parts of the source code can be reused by other case studies, covering other disciplines or technologies. 
When the results of our regularly scheduled case study will be evaluated and presented,
the analysis based on the first run will be re-evaluated and an in-depth interpretation of trends can be given.

If more case studies beyond the one presented in this paper will be implemented and executed,
two valuable achievements are in sight:
On the one hand, the results will help to gain a more complete picture of the research data landscape.
On the other hand, getting a higher score in most of the use-case-centric FAIR metrics will motivate providers of RDRs to be more FAIR.
It will also help researchers to use RDRs in a way that hopefully results in newly gained knowledge.

\section*{Acknowledgments}
We like to thank Stephan Hachinger and Tobias Guggemos for insightful feedback to prior versions of this paper.

The paper could not have been written without a lot of open source software.
Nevertheless the R package "knitr" (presented in \cite{rep016}) with its R integration into \LaTeX \  deserves specific praise for its outstanding usefulness in dynamically integrating statistical information.

This work was supported by the DFG (German Research Foundation) with the GeRDI project (Grants No. BO818/16-1).


\bibliographystyle{IEEEtran}
\bibliography{paper}
\end{document}